# Social Influence in the Concurrent Diffusion of Information and Behaviors in Online Social Networks

Kang Zhao[a1], Shiyao Wang[b], Ion B. Vasi[c], and Qi Zhang[d]

[a]Department of Management Sciences, University of Iowa, S224 PBB, Iowa City, IA 52242, USA.
[b]Department of Computer Science, University of Iowa, 14 MacLean Hall, Iowa City, IA 52242, USA.
[c]Department of Sociology, University of Iowa, W120 Seashore Hall, Iowa City, IA 52242, USA.
[d]School of Computer Science, Fudan University, 413-4 CS Bldg, 825 Zhangheng Rd, Shanghai, 201203 China.

**Abstract**

The emergence of online social networks has greatly facilitated the diffusions of information and behaviors. While the two diffusion processes are often intertwined, "talking the talk" does not necessarily mean "walking the talk"—those who share information about an action may not actually participate in it. We do not know if the diffusion of information and behaviors are similar, or if social influence plays an equally important role in these processes. Integrating text mining, social network analyses, and survival analysis, this research examines the concurrent spread of information and behaviors related to the Ice Bucket Challenge on Twitter. We show that the two processes follow different patterns. Unilateral social influence contributes to the diffusion of information, but not to the diffusion of behaviors; bilateral influence conveyed via the communication process is a significant and positive predictor of the diffusion of behaviors, but not of information. These results have implications for theories of social influence, social networks, and contagion.

---

[1] To whom correspondence should be addressed. E-mail: kang-zhao@uiowa.edu.

# Social Influence in the Concurrent Diffusion of Information and Behaviors in Online Social Networks

**Abstract**

The emergence of online social networks has greatly facilitated the diffusions of information and behaviors. While the two diffusion processes are often intertwined, "talking the talk" does not necessarily mean "walking the talk"—those who share information about an action may not actually participate in it. We do not know if the diffusion of information and behaviors are similar, or if social influence plays an equally important role in these processes. Integrating text mining, social network analyses, and survival analysis, this research examines the concurrent spread of information and behaviors related to the Ice Bucket Challenge on Twitter. We show that the two processes follow different patterns. Unilateral social influence contributes to the diffusion of information, but not to the diffusion of behaviors; bilateral influence conveyed via the communication process is a significant and positive predictor of the diffusion of behaviors, but not of information. These results have implications for theories of social influence, social networks, and contagion.

**Keywords**: social networks, diffusion, social influence, hazard model, machine learning,

## 1. Introduction.

The question of why some ideas and products generate interest by word of mouth, and grow virally, and why others don't, has received growing attention in the Age of the Internet. While some authors have argued that social epidemics are driven by "influentials" (Gladwell 2002), others have searched for the attributes that make ideas "contagious" –for example, their ability to possess social currency, or to be triggered, emotional, public, practical, and narrated (Berger 2016). Still others have examined the importance of social networks and, in particular, structural virality (Goel et al. 2015), tie strength and structural embeddedness (Aral and Walker 2014). Indeed, the diffusion of innovations and epidemics has been important topics for social network scholars (Eubank et al., 2004; Valente, 1995). Some studies have focused on the spread of information, commonly referred to as the "word of mouth" (Bakshy, Rosenn, Marlow, & Adamic, 2012; Leskovec, Backstrom, & Kleinberg, 2009; D. Watts & Dodds, 2007), while others have examined the spread of behaviors (Aral & Nicolaides, 2017; Aral & Walker, 2012; Bond et al., 2012; Centola, 2010; Christakis & Fowler, 2007).

The emergence of online social networks (OSNs), which have millions of users around the world, greatly facilitates the spread of information. OSNs have made the dissemination of information almost effortless. For example, OSN users can receive automatic updates of shared information (e.g., "News Feed" in Facebook and "Timeline" in Twitter) from their network neighbors, and then spread such information to his/her social network neighbors with a few clicks on a computer or mobile device (retweeting a tweet, sharing a status update, etc.). However, "talking the talk" in an OSN does not necessarily mean "walking the talk"; in other words, individuals may only share information about a behavior without actually adopting it. This is because behavioral decisions are shaped not only by the availability of information but also by factors such as personal commitment and resources, particularly for behaviors that occur offline (casting a vote at a polling station; participating in a demonstration; etc.). Research shows, for example, that the number of social media sites dedicated to the Occupy Wall Street movement was higher than the number of actual occupations, and that more people talked about, rather than participated in, this movement (Vasi & Suh, 2016).

Information and behaviors are often intertwined and diffuse concurrently. For instance, protest participants may disseminate information about a social movement's grievances, tactics, and collective actions before or at the same time as they protest. However, previous studies have investigated the diffusion of information and behaviors separately, making it very difficult to compare the two diffusion processes. Additionally, research shows that the likelihood that an individual adopts a particular behavior is strongly correlated with the behavior of his/her network neighbors. Such effects have been reflected in different diffusion models, such as threshold models and independent cascade models (Goldenberg, Libai, & Muller, 2001; Granovetter, 1978; D. J. Watts, 2002). Individuals may be influenced not only by peers and friends, but also by celebrities (Aral, 2013). Yet, we do not know if individuals are more likely to adopt a behavior when the celebrities they follow adopt the behavior, or when their friends have tried to persuade them. Consequently, the *first* goal of this research is to compare the diffusion of information and behaviors associated with a viral event; the *second* goal is to reveal the role of social influence from different social ties and via different processes for the diffusion of information and behaviors.

Using the Ice Bucket Challenge (IBC) as a case study and machine learning techniques, our research first identifies individuals' actual behaviors. This identification allows us to compare the diffusion patterns of information and behaviors related to the IBC from temporal, geospatial, individual, and network perspectives. Next, we use hazard models to study the roles of different types of social influence from different sources in the diffusion of information and behaviors. We conclude by discussing the main implications of this research for understanding who are the influential users are in OSNs, and how to develop more successful campaigns aiming at behavioral changes, such as charity donations, product adoptions, political mobilizations, and healthy lifestyles. In addition, our technique for detecting individuals' self-reported behaviors from publicly available text data also has the potential to enable more behavior diffusion research in large-scale social networks.

## 2. Theoretical background

Early research on diffusion focused mainly on the overall volume of "infections" (Hethcote, 1989) or product adoptions over time (Bass, 1969; Coleman, Katz, & Menzel, 1957). The emergence of information technologies and online social networks has made available fine-grained data on social ties, inter-personal interactions and individual decisions. Such data has enabled investigations of diffusions at dyadic and individual levels in email (Wu, Huberman, Adamic, & Tyler, 2004), blogging (Adar & Adamic, 2005), online gaming (Bakshy, Karrer, & Adamic, 2009), and social media (Goel, Anderson, Hofman, & Watts, 2015). While most of these studies focused on the diffusion of information, some have explored the spread of behaviors, such as voting (Bond et al., 2012), installing applications (Aral & Walker, 2014), using mobile services (Fang, Hu, Li, & Tsai, 2013), and purchasing online services ( Bapna & Umyarov, 2015). However, to our knowledge, no studies have investigated if the diffusion of behaviors follows a similar pattern as the diffusion of information.

In social networks, diffusion can be influenced by several factors. For example, people connected in a social network may get the same external stimuli from advertising campaigns and news coverages. Alternatively, diffusion can be shaped by homophily (a.k.a., "birds of a feather") (Aral, Muchnik, & Sundararajan, 2009; McPherson, Smith-Lovin, & Cook, 2001)--the phenomenon that describes how similar individuals tend to connected to each other in a social network. Homophily may lead to similar tastes or preferences between those who are connected by social ties and, thus, to interconnected decisions. Another important driver of diffusion is social influence (Aral & Nicolaides, 2017; Zhao et al., 2014), which happens when "an actor adapts his behavior, attitude, or belief, to the behaviors, attitudes, or beliefs of other actors in the social system" (Leenders, 2002). Although some scholars have argued that its impact on diffusions may be overestimated (Aral et al., 2009), the prevalence of online social networks offers new opportunities to leverage social influence to spread information and change behaviors.

Depending on who "other actors" are, social influence could come from two different sources, and may occur through different processes. First, an individual can be influenced by her acquaintances or friends –which is often described as peer influence (Marsden & Friedkin, 1993). Such influence is usually bilateral, because friends or acquaintances can often influence each other (the green arrows in Figure 1). Peer influence can occur through two processes: communication and/or comparison. *Communication* describes influence through direct contacts between the ego (the one who is influenced) and the alter (the one who influences others) (Homans, 1974). For instance, one decides to buy a type of smartphone after a friend recommended it to her. By contrast, *comparison* is indirect in nature--the ego uses alters as a reference for her own behaviors or opinions (Bem, 1967). An example would be one's purchase of a type of smartphone after seeing her friends doing so, even though none of them tried to persuade her to buy it.

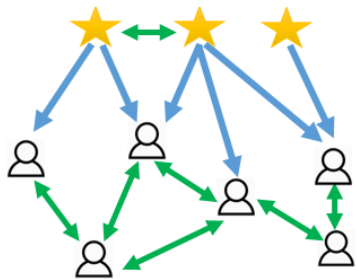

**Figure 1. An illustration of social influence.**

Second, individuals can be influenced by celebrities (Agrawal & Kamakura, 1995). Such influence is usually unilateral--from celebrities to "average Joes" (the blue arrows in Figure 1). For example, a fan may pay special attention to a celebrity's clothes or smartphone; yet, unless the

*6*

celebrity is endorsing those products, he/she is likely to be unaware of, and uninterested in her fans' consumer choices.

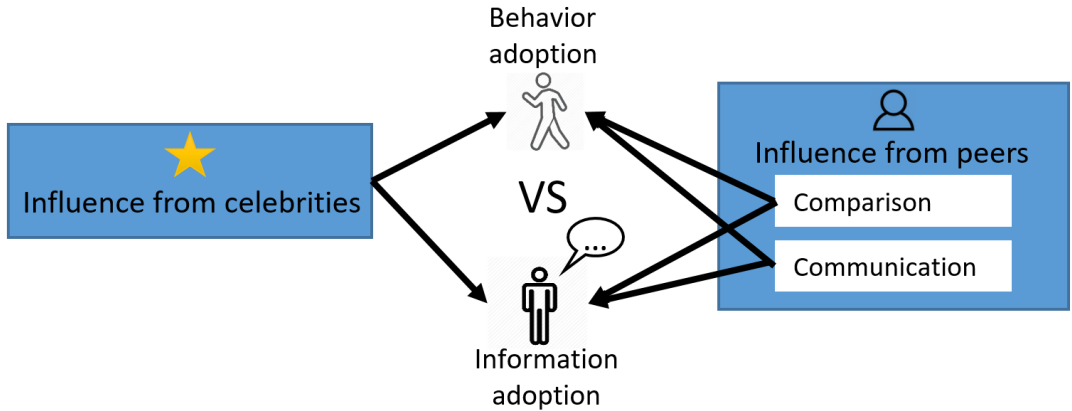

**Figure 2. A framework for analyzing social influence.**

Is social influence from peers more important than that from celebrities for both the diffusion of information *and* behaviors? To answer this question we need to compare the two processes directly. The diffusion of information and the diffusion of behaviors may be incommensurable if they examine different viral campaigns. Thus, we focus on the concurrent diffusion of information and behavior for the same viral event. We analyze the diffusion of information and behaviors associated with the Ice Bucket Challenge, because it was a major viral event that featured the concurrent diffusions of both information (e.g., talking about the IBC, sharing videos of celebrities' IBC) and behaviors (i.e., taking the IBC by pouring iced water over heads, or donating money to ALS-related agencies). The next section describes the dataset and the method used to identify behavior adoptions from IBC-related tweets.

## 3. The Identification of Behaviors

The IBC started in 2013 but was not associated with the amyotrophic lateral sclerosis (ALS) disease until June 30[th] 2014. After Peter Frates (an ALS survivor) nominated himself for the IBC on July 31[st], IBC became a viral campaign that ultimately raised tremendous amounts of money and awareness for ALS; for example, the IBC has been the fifth most popular Google search for all of 2014 and ALS has received in eight weeks thirteen times as much in contributions as what it had in the whole of the preceding year (Surowiecki, n.d.).

We used data from Twitter because Twitter is a major OSN that helped the success of the IBC. Our data was obtained from Twitter's firehose API using a set of keywords, phrases, and hashtags related to the IBC (e.g., variations of "ice bucket" and "#beatALS"); we collected IBC-related tweets that were written in English and posted between July 15$^{th}$ and Sep 15$^{th}$, 2014. The final dataset consists of 13.95 million tweets from 5.56 million users. There are 5.44 million original tweets (i.e., non-retweets), contributed by 2.5 million users; the remaining 8.51 million are retweets. The distribution of the number of IBC tweets per user in Figure 2 shows that most users had few IBC-related tweets: 60% of all the users in our dataset had only one IBC tweet, and 90% had fewer than 4.

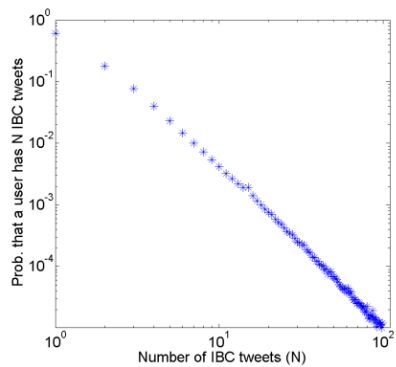

**Figure 3. The distribution of the number of IBC-related tweets per user.**

The diffusion of IBC information can be tracked by focusing on tweets with IBC-related keywords. To study the diffusion of behaviors associated with the IBC, we identified whether individuals stated that they took the IBC by examining their IBC-related messages. We used machine learning techniques to detect IBC behaviors because the scale of the dataset made human identifications of such behaviors impractical. The goal of the behavior detection was to classify each IBC tweet as either "*behavior tweets*", if the author took the IBC, or "*information tweets*", if the author did not take it. In other words, the detection was a text classification problem. Our behavior detection was limited to original tweets in the dataset because a retweet is a reference to someone else's IBC-related information or behavior. To train a text classifier, we hired Amazon Mechanical Turk (AMT) classification masters and asked them to read and annotate a random set of 7,000 original tweets to decide if a tweet offered enough evidence to conclude that its author took the Challenge. Each tweet was assigned to two AMT coders who passed our qualification tests; the Cohen's Kappa was 0.72. When they disagreed with each other on a tweet, a researcher from our team read the tweet and broke the tie.

Our feature set for text classification includes unigram and bigram vectors weighted by their TF-IDF scores, destinations of URLs embedded in tweets (e.g., whether it leads to a social networking website, a video/image sharing website, a news website, or others), the existence of related hashtags (e.g, #ALSchallenge), and the existence of numbers starting with monetary symbols (e.g., $100). We split the 7,000 annotated tweets into two sets—6,000 for 10-fold cross validation (CV) and 1,000 for hold-out testing. We tried several classification algorithms, and measured the performance of a classifier using 3 metrics.

1) Accuracy is the overall measure of how accurate a classifier is. It is defined as the total number of correctly classified instances (i.e., true positive and true negative) divided by the total number of instances.

2) F1 score is the harmonic mean of precision and recall. $F_1 = 2 * \frac{precision * recall}{precision + recall}$. The F1 score listed in Table 1 is for the positive class (i.e., behavior tweets).

3) AUC (area under the ROC curve) is the total area under the Receiver-Operating Characteristic (ROC) curve, which plots false positive rate vs true positive rate. It is a more robust measure of classifiers' performance when the prior distribution of the positive and negative classes is unbalanced. AUC=0.5 suggests a random classifier, while AUC=1 means a perfect classifier.

**Table 1. Tweet classification results from various classifiers**

|          | Decision tree (J48) | | SVM (linear kernel) | | Logistic Regression | | SVM (RBF kernel) | |
|----------|------|----------|------|----------|------|----------|------|----------|
|          | CV   | Hold-out | CV   | Hold-out | CV   | Hold-out | CV   | Hold-out |
| Accuracy | 0.878 | 0.858 | 0.899 | 0.884 | 0.897 | 0.886 | 0.922 | 0.908 |
| F1 score | 0.87 | 0.842 | 0.885 | 0.861 | 0.887 | 0.87 | 0.919 | 0.902 |
| AUC      | 0.777 | 0.699 | 0.723 | 0.665 | 0.898 | 0.864 | 0.955 | 0.941 |

After comparing the performance of top classifiers in Table 1, we picked SVM with the RBF kernel to classify all the other original tweets in our dataset. Among the 5.44 million original tweets, 22% were classified as the positive class or behavior tweets, since there was enough evidence that their authors took the IBC, whereas those classified as the negative class are information tweets. All retweets were automatically labeled as information tweets. If a user posted at least one behavior tweet, then he/she is considered an IBC **doer**; those who posted only information tweets are considered IBC **talkers**.

## 4. Diffusion Patterns
### *4.1. Temporal and geospatial patters*
Figure 4 shows the percentage of information tweets (including both original tweets and retweets) and behavior tweets for each day during the 2-month period of the study. Both curves are almost flat before Aug 12$^{th}$, 2014 but show great increases between Aug 15$^{th}$ and 19$^{th}$, when many celebrities took the IBC. For example, Bill Gates took the IBC on Aug 15$^{th}$, Taylor Swift on the 16$^{th}$, Justin Bieber on the 17$^{th}$, Lady Gaga on the 18$^{th}$, and Katy Perry on the 19$^{th}$. The increase for information tweets is much faster than that for behavior tweets; both reached peaks on Aug 21$^{st}$, with information tweets having a higher peak. However, right before the peak, the information tweets curve takes a major dip on Aug 20th, while behavior tweets decreased only slightly. We conjecture that the dip was caused by other events that distracted people's attention from the IBC. For example, American journalist James Foley was beheaded on the 19th and U.S. Attorney General Eric Holder visited Fergusson, Missouri on the 20th. After the peak, the information tweets curve also drops faster than the behavior tweets curve.

Temporal dynamics of information and behavior diffusions show a few differences. The volume of IBC-related information features exponential growth to a high peak and exponential decay after that, which is similar to the temporal trends of other viral stories in online social networks. We argue that this is the result of the fact that, while the rapid spread of information in OSNs is almost effortless, the continued diffusion of information requires sustained attention, which is difficult to maintain in the Information Age. By contrast, the trend of IBC behaviors has lower spikes and a slower decay. This is most likely because adopting a non-trivial behavior, such as taking the IBC, requires substantial planning (for example, finding a location and preparing for the Challenge).

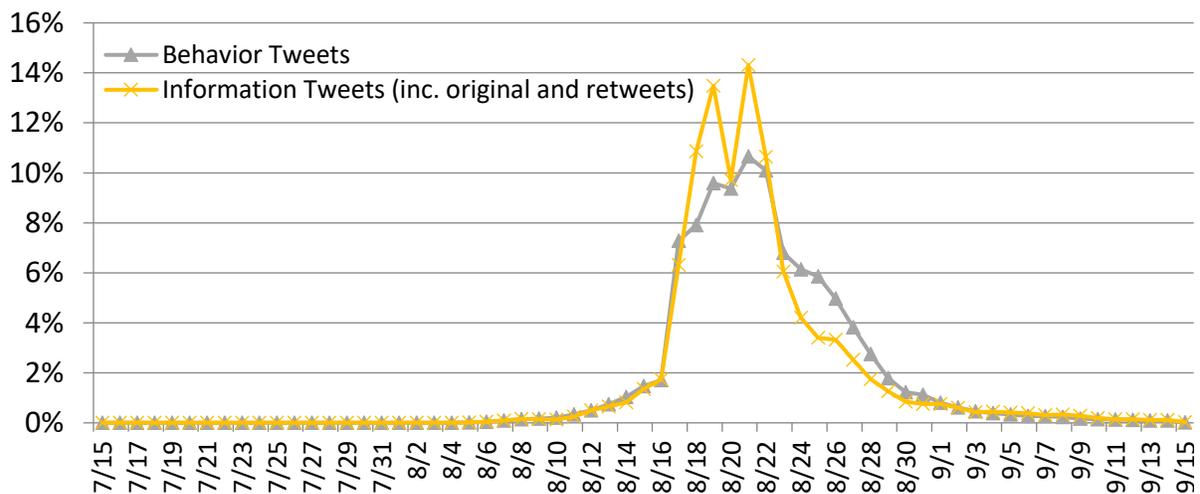

**Figure 4. The percentage of IBC information and behavior tweets over time.**

We also retrieved geographical locations of users who had IBC tweets from their Twitter profiles. The retrieval was limited to users whose tweets have the language code "en" (i.e., English). For users who provided state information, we directly matched the state with full names or abbreviations of 50 states of the U.S. and Washington D.C. Abbreviations of a state include both the two-letter ANSI code (e.g., TX for Texas and CA for California), and the Associated Press Abbreviation (e.g., Fla. for Florida, and Mass. for Massachusetts). For those who provided only a city, we used the list of top 50 cities in the U.S. by population, and assigned a matched city to the corresponding U.S. states. Through such a procedure, we identified 1.99 million users from 50 U.S. states and Washington DC.

We then calculated two measures for each state: (1) information adoption rate, which is the ratio between the number of users who have an IBC tweet in a state and the number of the state's Internet users (estimated by the state population multiplied by the state's Internet penetration rate), and (2) behavior conversion rate, which is the ratio between the number of IBC doers and the number of users who have IBC tweets (i.e., talkers and doers). Among all the states, information adoption rates range from 0.2% to 4.4%; while behavior conversation rates have a minimum value of 10% and a maximum value of 14.2%. Figure 5 shows a map of U.S. with the two rates for each state. Table 2 lists top 5 and bottom 5 states by the two measures. Only one state, Kansas, appears in the top 5 of both measures, and bottom 5 states of the two measures do not overlap at all. Moreover, the Spearman's rank correlation coefficient between states' ranks by both measures is only 0.168.

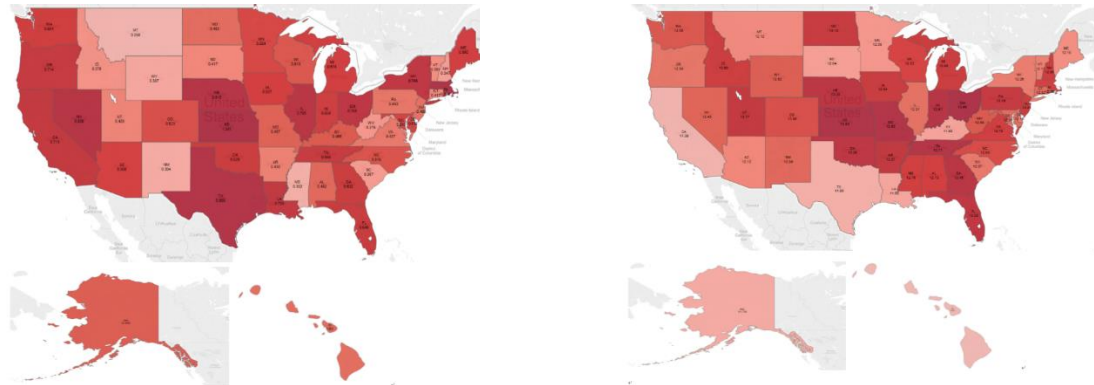

    **(a) Information adoption rates**        **(b) Behavior conversion rates**

**Figure 5. The information adoption rates and behavior conversion rates for U.S. states. Darker colors mean higher values.**

**Table 2. Top and bottom states by information adoption rates and behavior conversion rates.**

|  | 1st | 2nd | 3rd | 4th | 5th |
|---|---|---|---|---|---|
| Top 5 states by information adoption rates | DC | KS | MA | TX | NE |
| Top 5 states by behavior conversion rates | OH | KS | MO | TN | IN |
| Bottom 5 states by information adoption rates | MT | MS | NM | WY | NH |
| Bottom 5 states by behavior conversion rates | HI | CA | TX | AK | LA |

This again indicates that there is a difference between the diffusion of information and behaviors: a state with a high number of IBC talkers does not necessarily have a high number of IBC doers. An example is the state of California— it is the 12th in information adoption rate, but second to last among all states in behavior conversion rate. We hypothesize that this result may be related to concerns about wasting water by taking the IBC during the state's record-breaking drought in 2014.

### *4.2. Network diffusion patterns*

To find evidence of diffusion via social networks, we separated individual adoptions in two types: *spontaneous* and *viral* adoptions (Aral & Walker, 2012). In the context of Twitter, a spontaneous adoption refers to the adoption of IBC information or behavior before the adopter was exposed to IBC-related information from his/her immediate followees (i.e., those followed by a user). Conversely, if a user tweeted about the IBC after one of her followees tweeted about IBC, or took the IBC after one of her followee took it, the adoption is considered viral. It is important to note that the name "viral adoption" does not necessarily mean the user's adoption was caused by her followees' adoption.

We chose a random sample of 20,000 IBC talkers and 10,000 IBC doers because talkers outnumber doers. We retrieved their Twitter followees, identified talkers or doers, and checked the time of the IBC information and behavior tweets. We found that viral adoptions dominate both information and behavior adoptions: 99.1% of the 20,000 IBC talkers posted their first IBC tweets after at least one of their followees posted IBC tweets. The followees who talked about the IBC (without taking it) before a target talker's first IBC tweet are referred to as "*followee talkers*". Similarly, 96.4% of the 10,000 IBC doers took the IBC after at least one or more of their followees took it; these are called "*followee doers*".

As most IBC information and behavior adoptions are viral adoptions, we examined adoption rates controlling for the number of social network neighbors who have adopted. In other words, such a rate is the conditional probability that an individual adopts information or behavior given that a certain number of her followees have adopted. The calculation was based on a random sample of 31,500 Twitter users who registered before July 2014 and posted tweets in English[2]. Among these users, we identified 1,489 talkers and 324 doers. The "Followee talkers vs Information adoption" curve in Figure 6 plots the cumulative probability that a user adopted IBC information given the number of followee talkers, and the "Followee doers vs Behavior adoption" curve shows cumulative probability that a user took the IBC given the number of followee doers. Both distributions resemble logarithmic growth curves, which suggests that having more talkers (or doers) in one's social network neighborhood is generally associated with higher chances of information (or behavior) adoption, although the marginal return gradually diminishes, especially for behavior adoptions. As we would expect, the adoption rate is higher for information than for behavior, controlling for the number of followee adopters. For example, the information adoption rate reaches 2.77% for those with fewer than or equal to 100 followee talkers. By contrast, the behavior adoption rate for those with fewer than or equal to 100 followee doers is only 0.92%. This is consistent with our assumption that sharing information in an OSN is easier than adopting a behavior.

---

[2] We randomly sampled 50,000 Twitter user IDs, but only 31,500 had valid profiles

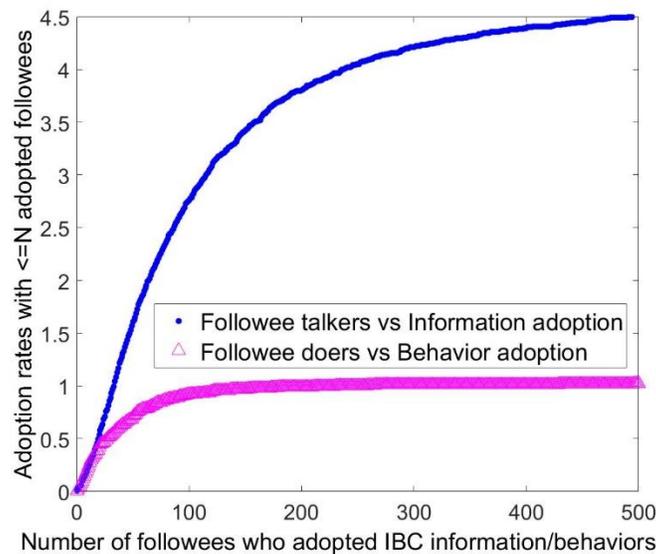

**Figure 6. Cumulative distribution of adoption rates for a random sample of 31.5k users.**

## 5. Untangling Social Influence

To investigate the effects of various types of social influence on the diffusion of IBC information and self-reported behaviors, we used survival analysis. More specifically, we used two continuous-time Cox proportional hazard models, an approach widely used to study social influence (Aral & Walker, 2012): the first one is for information adoption, with the dependent variable being the time to individuals' first IBC information tweet; the second one is for behavior adoption, with the dependent variable time to individuals' IBC behaviors. Table 3 summarizes control and independent variables in the two models.

**Table 3.** Variables used for hazard models.

|  | Model-1: Information adoption | Model-2: Behavior adoption |
|---|---|---|
| Dependent var. | Time to information adoption | Time to behavior adoption |
| Control variables | A user's seniority in Twitter (*Seniority*) | |
| | A user's average number of tweets per month (*TweetPerMon*) | |
| | The percentage of retweets for a user (*%Retweets*) | |
| | A user's number of followers (*#Followers*) | |
| | A user's number of followees (*#Followees*) | |
| | Topic distribution of a user's tweets | |
| Independent variables | The number of IBC info. tweets from a user's unilateral followees (*UniInfo*) | The number of IBC behavior tweets from a user's unilateral followees (*UniBeh*) |
| | The number of IBC info. tweets from a user's bilateral friends (*BiInfoComp*) | The number of IBC behavior tweets from a user's bilateral friends (*BiBehComp*) |
| | The number of IBC info. tweets mentioning a user from the user's bilateral friends (*BiInfoComm*) | The number of IBC behavior tweets mentioning a user from the user's bilateral friends (*BiBehComm*) |

As the goal is to study the effects of social influence, independent variables for the models capture social influence from a user's social network neighbors. In the information adoption model (Model 1), social influence is based on users' exposure to IBC information from her Twitter followees; in the behavior adoption model, social influence is based on users' exposure to IBC behavior from her followees.

Based on the framework in Figure 2, we modeled two sources of social influence from an individual's Twitter followees: unilateral followees and bilateral followees. First, unilateral followees are those who are followed by a user but do not follow the user back –these are often celebrities whom a user is interested in following. Second, bilateral followees are those who have a bilateral following relationship with the user. They are usually the user's friends, family or peers.

As we described earlier, peer influence can occur via two processes. When a peer explicitly mention the user mentions a user (via @) in her tweet, the communication process is at work. For example, a friend posted an information tweet that mentioned the user to draw her attention to IBC--"@JohnDoe, you should check out Bill Gate's Ice Bucket Challenge". It could also be a behavior tweet that challenges or nominates others (e.g., "Here is my Ice Bucket Challenge video .... Now I challenge @MarySmith"). By contrast, peer influence occurs via the comparison process when peers' IBC tweets did not mention the user. All these variables are measured by the number of information or behavior adopters in an individuals' social network neighborhood till time *t*.

We also tried to control for confounding factors. First, we included several control variables that reflect users' individual characteristics, because such a viral campaign may only attract Twitter users with certain demographics. We included how active a user is (e.g., seniority and the number of tweets per month), a user's retweeting patterns (e.g., %Retweets), and a user's social network centralities (e.g., the numbers of followers and followees). However, these can hardly reflect what a user's interests are, and do not take advantages of the rich source of text data from tweets. Therefore, we also applied author-topic models (Rosen-Zvi, Griffiths, Steyvers, & Smyth, 2004) to non-IBC tweets collected from users' Twitter timelines. In the results of classic topic models, such as LDA (Blei, 2012), each topic is represented by a list of keywords and each document has a distribution over these topics. Author-topic model extends topic models by providing an author-topic distribution to show probabilities that a user has covered these topics in documents (i.e., tweets in this case) published by the user. In other words, we generated a topic distribution for each user to represent her overall topical interests on Twitter. For example, a Twitter user with a topic distribution <0.1, 0.05, 0.85> would mean the user's tweets focus more on the $3^{rd}$ topic, instead of the $1^{st}$ and the $2^{nd}$. Further, by using continuous-time hazard models which compares users' adoptions during the same period, our model can also tease out some effects of external influence outside one's social network neighbors (e.g., from mass media) if we assume individuals receive similar amount of external influence from the mass media at the same time.

To ensure the robustness of the results, we picked two random samples of users for hazard modeling. Both samples initially include 10,000 randomly selected user IDs. After removing those who have protected or invalid accounts, the 1st sample has 6,117 users, including 1,471 talkers and 332 doers, while the 2nd has 6,168 users, including 1,388 talkers and 312 doers. Continuous-time hazard models were constructed for each user on hourly basis from 8/15/2014 to 8/29/2014. Covering one week before and after the peak day (8/22/2014) for the IBC, this period represents the time when the event was popular. Those who did not adopt IBC information or behaviors were right-censored. In our experiments, we collected all available tweets from the timelines for each user in the two random samples. Stop words common in English (e.g., "the" and "of"), as well as stop words popular in tweets (e.g., "RT", "http") were removed before analyzing the texts of tweets. Table 4 shows keywords for each topic from tweets in our experiments. Topics are listed in the descending order of prevalence.

Table 5 lists hazard ratios from hazard models for the two samples; a ratio greater than 1 means higher values of the variable are associated with higher chances of adoptions, and vice versa. For example, the hazard ratio 1.105 for seniority in Sample-1 for Model-1 means that an increase of one standard deviation in seniority is associated with 10.5% increase in the probability of adopting IBC-related information.

**Table 4. Topics generated by author topics models (k=20).**

| Topic | Top 5 keywords |
|---|---|
| 1 | Obama, president, American, muslim, gun |
| 2 | business, job, market, manager, social |
| 3 | people, feel, care, hate, time |
| 4 | photo, post, tonight, travel, city |
| 5 | lol, fuck, shit, nigga, bitch |
| 6 | life, god, love, word, heart |
| 7 | come, free, enter, money, sale |
| 8 | game, play, team, win, fan |
| 9 | time, day, school, start, sleep |
| 10 | food, eat, drink, water, coffee |
| 11 | love, day, happy, thank, hope |
| 12 | story, look, book, read, art |
| 13 | watch, movie, film, star, world |
| 14 | follow, please, thank, via, unfollow |
| 15 | video, music, check, song, listen |
| 16 | love, girl, look, boy, justin |
| 17 | hahaha, aku, yang, ada, dan |
| 18 | que, por, para, san, del |
| 19 | yes, lol, haha, OK, till |
| 20 | van, een, meet, het, die |

Results show that the two user samples yielded consistent results for the effects of social influence (see Table 5). *First*, unilateral social influence from celebrities is associated with the diffusion of information, but not behaviors. In other words, when a user saw more IBC information tweets from celebrities he/she follows, the user was more likely to post an IBC information tweet. However, when more unilateral followees took IBC, the user is not significantly more likely to adopt the IBC behavior. *Second*, peer influence via comparison is positively correlated with both diffusion processes. If more of a user's bilateral friends talked about or took the IBC, the user is significantly more likely to be an IBC talker, or doer respectively. *Third*, peer influence via communication does not predict information diffusion, but is a significant and positive predictor of behavior adoption. It is interesting that, while a doer's direct challenging of another peer can increase the chance of behavior adoption by the one being challenged, a peer's explicit attempt to draw one's attention to IBC information does not result in information diffusion.

**Table 5.** Hazard ratios for the two hazard models (with standard errors in parentheses, and control variables in shade).

|  | Model-1: Information adoption | | | Model-2: Behavior adoption | |
| --- | --- | --- | --- | --- | --- |
|  | Sample-1 | Sample -2 |  | Sample -1 | Sample -2 |
| Seniority | 1.105**($1.1 \times 10^{-4}$) | 1.152***($1.4 \times 10^{-4}$) | Seniority | 1.079($1.3 \times 10^{-4}$) | 1.212*($1.4 \times 10^{-4}$) |
| TweetPerMonth | 2.119***($5.0 \times 10^{-4}$) | 1.844***($5.0 \times 10^{-4}$) | TweetPerMonth | 2.160***($4.7 \times 10^{-4}$) | 2.308***($4.7 \times 10^{-4}$) |
| %Retweets | 1.261***($1.9 \times 10^{-4}$) | 1.256***($1.9 \times 10^{-4}$) | %Retweets | 0.865*($1.8 \times 10^{-4}$) | 0.865*($1.8 \times 10^{-4}$) |
| # Followers | 0.950($4.5 \times 10^{-4}$) | 0.944($4.5 \times 10^{-4}$) | # Followers | 0.997($4.2 \times 10^{-4}$) | 1.070($4.2 \times 10^{-4}$) |
| #Followees | 0.766***($3.9 \times 10^{-4}$) | 0.825***($3.9 \times 10^{-4}$) | #Followees | 0.816*($3.6 \times 10^{-4}$) | 0.773*($3.6 \times 10^{-4}$) |
| Topics | 6 topics have p<0.05 | | Topics | 1 topic has p<0.05 | |
| UniInfo | 1.178***($4.3 \times 10^{-4}$) | 1.132***($6.1 \times 10^{-4}$) | UniBeh | 1.121($5.9 \times 10^{-4}$) | 1.178($5.9 \times 10^{-4}$) |
| BiInfoIndirect | 1.789***($5.0 \times 10^{-4}$) | 1.603***($6.1 \times 10^{-4}$) | BiBehIndirect | 1.743***($5.9 \times 10^{-4}$) | 1.945***($5.9 \times 10^{-4}$) |
| BiInfoDirect | 0.949($2.2 \times 10^{-5}$) | 1.013($2.1 \times 10^{-5}$) | BiBehDirect | 1.110***($3.8 \times 10^{-5}$) | 1.226***($3.4 \times 10^{-5}$) |
| *: p<0.05, **: p<0.01, ***: p<0.001 | | | | | |

As for control variables, individual characteristics also have a number of interesting effects. Overall, individuals' characteristics matter more for information diffusion than for behavior diffusion. More active Twitter users are more likely to become both IBC talkers and doers. Those who have higher percentages of retweets are more likely to talk about IBC, but have lower chances of taking the IBC. In addition, a user's topic of interests besides IBC is a better predictor for information adoptions (with distributions on 6 topics being significant) than for behavior adoptions (with only one topic being significant) during IBC. This is not surprising because a user's topical interests are reflected by her previous tweets, which represent her information sharing patterns. Talking about the IBC is essentially sharing information too. Consequently, compared to taking the Challenge, adopting IBC information should be better captured by one's previous information sharing.

In sum, the results untangled the roles of social influence in the concurrent diffusion of information and behavior. There are both similarities and differences. Social influence from peers in an OSN is associated with the adoption of both information and behaviors, when such influence is conveyed indirectly via the comparison process. The effect of peer influence conveyed via direct communications is also significant, suggesting that persuasion from a peer is an effective way to spread the behavior. These results lend support to viral marketing efforts, because information and behaviors do spread among friends or peers connected by bilateral social ties. However, the effect of influence from unilateral sources is limited to the diffusion of information. This finding has implications for viral campaigns that plan to leverage large fan bases of famous individuals: even though celebrities can start the "word of mouth", their actions may have little impact on other people's actions.

## 6. Discussion

Combining machine learning, network analysis, and survival analysis, this study compared the diffusion processes of information and behaviors in online social networks. Based on large-scale data about the Ice Bucket Challenge from Twitter, we identified individuals' actual behaviors with regard to the Challenge, and distinguished those who took IBC from those who only talked about IBC on Twitter. Comparing the diffusion patterns of IBC information and behaviors head-to-head, we revealed similarities and differences from temporal, geospatial, and network perspectives. Based on a framework of social influence for information and behavior diffusions, survival analyses found that social influence from different sources and conveyed in different processes could play different roles in the concurrent diffusion of information and behaviors for IBC.

This study has a number of implications for the study of OSNs. First, this research demonstrates the possibility of identifying self-reported offline behaviors using computational methods. As the IBC case shows, only a relatively small proportion of those who talk about an event engage in non-trivial actions related to the event. Our research shows that carefully designed machine learning algorithms can be used for behavior identification. Second, activists, viral campaign organizers, and marketers should think twice about the effectiveness of using so-called "influencers" in OSNs to influence the adoption of offline behaviors. Influencers with large fan bases may be able to spread information (e.g., via retweets), but their ability to influence people's offline behaviors is limited. Despite a common perception that celebrities have a powerful influence in the Age of Social Media, our study shows that for most people the adoption of a behavior is shaped more by the actions of their peers than by those of celebrities. A promising direction for future research is to use OSNs to examine the effect of various types of influencers on the adoption of different behaviors, so that we can better define and identify influential users in OSNs.

This study, however, also has limitations. One limitation is that we rely only on Twitter data to analyze diffusion processes. We recognize the possibility that some individuals have reported the adoption of IBC information or behaviors in other OSNs, or did not report it on OSNs at all. Even though we controlled for as many individual characteristics as we could, we were not able to control for individuals' use of other OSNs platforms, or to observe their actual behaviors. Future diffusion studies that incorporate data from multiple OSNs platforms, and consolidate users across different platforms (Vosecky, Hong, & Shen, 2009), could help to generalize our findings. Another limitation is that our detection of individual behaviors is based on self-reported behaviors, not observed behaviors. Future studies that corroborate the association between self-reported behaviors and behaviors that are observed (for example, from available pictures or video) could advance understanding of the relationship between the diffusion of information and both self-reported and observed behaviors. Third, even though we controlled for several confounding factors in our model, it is inaccurate to claim that the relationship between social influence and behavior adoption is causal. Because running randomized experiments for such viral events can be challenging, better statistical methods (e.g., propensity score matching or instrument variables) are needed to investigate causality.

# References

Adar, E., & Adamic, L. A. (2005). Tracking Information Epidemics in Blogspace. In *Proceedings of the 2005 IEEE/WIC/ACM International Conference on Web Intelligence* (pp. 207–214). Washington, DC, USA: IEEE Computer Society. https://doi.org/10.1109/WI.2005.151

Agrawal, J., & Kamakura, W. A. (1995). The Economic Worth of Celebrity Endorsers: An Event Study Analysis. *Journal of Marketing*, *59*(3), 56–62. https://doi.org/10.2307/1252119

Aral, S. (2013). What Would Ashton Do—and Does It Matter? *Harvard Business Review*, (May). Retrieved from https://hbr.org/2013/05/what-would-ashton-do-and-does-it-matter

Aral, S., Muchnik, L., & Sundararajan, A. (2009). Distinguishing influence-based contagion from homophily-driven diffusion in dynamic networks. *Proceedings of the National Academy of Sciences*, *106*, 21544–21549. https://doi.org/10.1073/pnas.0908800106

Aral, S., & Nicolaides, C. (2017). Exercise contagion in a global social network. *Nature Communications*, *8*, 14753. https://doi.org/10.1038/ncomms14753

Aral, S., & Walker, D. (2012). Identifying Influential and Susceptible Members of Social Networks. *Science*, *337*, 337–341. https://doi.org/10.1126/science.1215842

Aral, S., & Walker, D. (2014). Tie Strength, Embeddedness, and Social Influence: A Large-Scale Networked Experiment. *Management Science*, *60*(6), 1352–1370. https://doi.org/10.1287/mnsc.2014.1936

Bakshy, E., Karrer, B., & Adamic, L. A. (2009). Social Influence and the Diffusion of User-created Content. In *Proceedings of the 10th ACM Conference on Electronic Commerce* (pp. 325–334). New York, NY, USA: ACM. https://doi.org/10.1145/1566374.1566421

Bakshy, E., Rosenn, I., Marlow, C., & Adamic, L. (2012). The role of social networks in information diffusion. In *Proceedings of the 21st international conference on World Wide Web* (pp. 519–528). New York, NY, USA: ACM. https://doi.org/10.1145/2187836.2187907

Bapna, R., & Umyarov, A. (2015). Do Your Online Friends Make You Pay? A Randomized Field Experiment on Peer Influence in Online Social Networks. *Management Science.*, *61*(8), 1902–1920. https://doi.org/10.1287/mnsc.2014.2081

Bass, F. M. (1969). NEW PRODUCT GROWTH FOR MODEL CONSUMER DURABLES. *Management Science Series A-Theory*, *15*, 215–227.

Bem, D. J. (1967). Self-perception: An alternative interpretation of cognitive dissonance phenomena. *Psychological Review*, *74*(3), 183–200.

Blei, D. M. (2012). Probabilistic topic models. *Commun. ACM*, *55*(4), 77–84. https://doi.org/10.1145/2133806.2133826

Bond, R. M., Fariss, C. J., Jones, J. J., Kramer, A. D. I., Marlow, C., Settle, J. E., & Fowler, J. H. (2012). A 61-million-person experiment in social influence and political mobilization. *Nature*, *489*, 295–298.

Centola, D. (2010). The Spread of Behavior in an Online Social Network Experiment. *Science*, *329*, 1194–1197. https://doi.org/10.1126/science.1185231

Christakis, N. A., & Fowler, J. H. (2007). The Spread of Obesity in a Large Social Network over 32 Years. *New England Journal of Medicine*, *357*, 370–379. https://doi.org/10.1056/NEJMsa066082

Coleman, J., Katz, E., & Menzel, H. (1957). The Diffusion of an Innovation Among Physicians. *Sociometry*, *20*(4), 253–270. https://doi.org/10.2307/2785979

Eubank, S., Guclu, H., Anil Kumar, V. S., Marathe, M. V., Srinivasan, A., Toroczkai, Z., & Wang, N. (2004). Modelling disease outbreaks in realistic urban social networks. *Nature*, *429*(6988), 180–184. https://doi.org/10.1038/nature02541

Fang, X., Hu, P. J.-H., Li, Z. (Lionel), & Tsai, W. (2013). Predicting Adoption Probabilities in Social Networks. *Information Systems Research*, *24*(1), 128–145. https://doi.org/10.1287/isre.1120.0461

Goel, S., Anderson, A., Hofman, J., & Watts, D. J. (2015). The Structural Virality of Online Diffusion. *Management Science*, *62*(1), 180–196. https://doi.org/10.1287/mnsc.2015.2158

Goldenberg, J., Libai, B., & Muller, E. (2001). Talk of the network: A complex systems look at the underlying process of word-of-mouth. *Marketing Letters*, *12*, 211–223.

Granovetter, M. (1978). Threshold Models of Collective Behavior. *The American Journal of Sociology*, *83*, 1420–1443.

Hethcote, H. W. (1989). Three Basic Epidemiological Models. In S. A. Levin, T. G. Hallam, & L. J. Gross (Eds.), *Applied Mathematical Ecology* (pp. 119–144). Springer Berlin Heidelberg. https://doi.org/10.1007/978-3-642-61317-3_5

Homans, G. C. (1974). *Social Behavior: Its Elementary Forms*. Harcourt Brace.

Leenders, R. T. A. J. (2002). Modeling social influence through network autocorrelation: constructing the weight matrix. *Social Networks*, *24*(1), 21–47. https://doi.org/10.1016/S0378-8733(01)00049-1

Leskovec, J., Backstrom, L., & Kleinberg, J. (2009). Meme-tracking and the dynamics of the news cycle (pp. 497–506). Presented at the Proceedings of the 15th ACM SIGKDD international conference on Knowledge discovery and data mining, 1557077: ACM. https://doi.org/10.1145/1557019.1557077

Marsden, P. V., & Friedkin, N. E. (1993). Network Studies of Social Influence. *Sociological Methods & Research*, *22*(1), 127–151. https://doi.org/10.1177/0049124193022001006

McPherson, M., Smith-Lovin, L., & Cook, J. M. (2001). Birds of a Feather: Homophily in Social Networks. *Annual Review of Sociology*, *27*, 415–444. https://doi.org/10.1146/annurev.soc.27.1.415

Rosen-Zvi, M., Griffiths, T., Steyvers, M., & Smyth, P. (2004). The Author-topic Model for Authors and Documents. In *Proceedings of the 20th Conference on Uncertainty in Artificial Intelligence* (pp. 487–494). Arlington, Virginia, United States: AUAI Press. Retrieved from http://dl.acm.org/citation.cfm?id=1036843.1036902

Surowiecki, J. (n.d.). What Happened to the Ice Bucket Challenge? Retrieved July 26, 2016, from http://www.newyorker.com/magazine/2016/07/25/als-and-the-ice-bucket-challenge

Valente, T. W. (1995). *Network Models of the Diffusion of Innovations*. Cresskill, N.J: Hampton Press.

Vasi, I. B., & Suh, C. S. (2016). Online Activities, Spatial Proximity, and the Diffusion of the Occupy Wall Street Movement in the United States. *Mobilization: An International Quarterly*, *21*(2), 139–154. https://doi.org/10.17813/1086-671X-22-2-139

Vosecky, J., Hong, D., & Shen, V. Y. (2009). User identification across multiple social networks. In *2009 First International Conference on Networked Digital Technologies* (pp. 360–365). https://doi.org/10.1109/NDT.2009.5272173

Watts, D., & Dodds, P. (2007). Influentials, Networks, and Public Opinion Formation. *Journal of Consumer Research*, *34*, 441–458. https://doi.org/10.1086/518527

Watts, D. J. (2002). A simple model of global cascades on random networks. *Proceedings of the National Academy of Sciences of the United States of America*, *99*, 5766–5771.

Wu, F., Huberman, B. A., Adamic, L. A., & Tyler, J. R. (2004). Information flow in social groups. *Physica A-Statistical Mechanics and Its Applications*, *337*, 327–335. https://doi.org/10.1016/j.physa.2004.01.030

Zhao, K., Yen, J., Greer, G., Qiu, B., Mitra, P., & Portier, K. (2014). Finding Influential Users of Online Health Communities: a New Metric based on Sentiment Influence. *Journal of the American Medical Informatics Association*, *21*(e2), e212–e218. https://doi.org/10.1136/amiajnl-2013-002282